\documentclass[sigconf,nonacm]{acmart}

\setcopyright{none}
\renewcommand\footnotetextcopyrightpermission[1]{}
\usepackage{algorithm}
\usepackage{algpseudocode}
\usepackage{booktabs}
\usepackage{multirow}
\usepackage{makecell}
\usepackage{graphicx}
\usepackage{subcaption}
\usepackage{array}
\usepackage{tabularx}
\usepackage{placeins}
\usepackage{comment}
\usepackage{adjustbox}
\usepackage{microtype}

\AtBeginDocument{%
}

\begin{document}

\title{CHARMING: Parallel Edge AI Acceleration Across Tensor Processing
and In-Memory Computation}


\author{Md Omar Faruque}
\affiliation{%
  \institution{Syracuse University}
  \city{Syracuse}
  \state{New York}
  \country{USA}}
\email{mdfaruqu@syr.edu}

\author{Jingang Jin}
\affiliation{%
  \institution{Syracuse University}
  \city{Syracuse}
  \state{New York}
  \country{USA}}
\email{jjin24@syr.edu}

\author{Jo\~ao Paulo C. de Lima}
\affiliation{%
  \institution{Friedrich-Alexander-Universit\"at Erlangen-N\"urnberg}
  \city{Erlangen}
  \country{Germany}}
\email{joao.lima@fau.de}

\author{Shouzhi Fang}
\affiliation{%
  \institution{Syracuse University}
  \city{Syracuse}
  \state{New York}
  \country{USA}}
\email{sfang18@syr.edu}

\author{William C. Tegge}
\affiliation{%
  \institution{Syracuse University}
  \city{Syracuse}
  \state{New York}
  \country{USA}}
\email{wtegge@syr.edu}

\author{Peipei Zhou}
\affiliation{%
  \institution{Brown University}
  \city{Providence}
  \state{Rhode Island}
  \country{USA}}
\email{peipei\_zhou@brown.edu}

\author{Qinru Qiu}
\affiliation{%
  \institution{Syracuse University}
  \city{Syracuse}
  \state{New York}
  \country{USA}}
\email{qiqiu@syr.edu}

\author{Alex K. Jones}
\affiliation{%
  \institution{Syracuse University}
  \city{Syracuse}
  \state{New York}
  \country{USA}}
\email{akj@syr.edu}

\renewcommand{\shortauthors}{Faruque et al.}


\begin{abstract}
Edge AI increasingly relies on specialized accelerators for low-latency
inference under tight resource and energy constraints, yet their efficiency
remains limited by data movement. Prior work, CHARM, mitigates this overhead
by overlapping data movement with computation. We observe that in-memory
computing can further supplement tensor processing as a parallel accelerator
while eliminating certain data movements entirely.

We present CHARMING, a heterogeneous edge AI architecture that uses
accuracy- and quantization-aware partitioning to distribute and concurrently
execute inference workloads across tensor processing and in-memory
computation. CHARMING partitions attention heads between an AIE array and an
in-DRAM compute unit, balancing execution according to model architecture,
sequence length, and precision while selecting heads to minimize accuracy
degradation. CHARMING also integrates quantization into the AIE projection
stage to further reduce data movement. We validate CHARMING end-to-end in
hardware on AMD Versal VCK5000 and Alveo U200 platforms and evaluate LLM and
video-generation models across multiple precisions using measured hardware
results and simulation of the latest in-memory compute hardware.

CHARMING improves prefill performance by up to 23.6\% and decode performance
by up to 44\% over AIE-only acceleration. With
\(\mathrm{int4}\times\mathrm{int4}\) fixed-point in-memory computation,
CHARMING reduces per-layer energy by up to 7.6\% and improves energy-delay
product by up to 24.7\%, while maintaining accuracy similar to a 32-bit
floating-point baseline.
\end{abstract}




\maketitle
\pagestyle{plain}

\section{Introduction}

The growing computational demands of artificial intelligence have motivated a broad range of specialized hardware. Graphics processing units (GPUs) remain the dominant platform for deep learning because their massive thread-level parallelism provides substantially higher throughput than general-purpose CPUs on data-parallel workloads~\cite{brodtkorb2013graphics,dally2021evolution}. Recent GPUs further improve efficiency through dedicated matrix engines, such as Tensor Cores, and reduced-precision arithmetic~\cite{markidis2018nvidia,choquette2021nvidia}. 
Nevertheless, deploying these accelerators at the edge remains challenging under tight memory, energy, and computational resource budgets.

\begin{sloppypar}
\looseness=-1
Edge devices increasingly execute inference locally to reduce communication latency and cloud-computing costs, preserve data privacy, and operate without continuous network connectivity~\cite{qu2025mobile,mao2024green}. 
At the same time, edge workloads are evolving from compact convolutional networks to transformer-based language and multimodal models whose storage, computation, and memory-bandwidth requirements can exceed device capabilities~\cite{zheng2025review,zhou2024survey}. 
Dense self-attention scales quadratically with sequence length, while autoregressive decoding repeatedly accesses model weights and a growing key--value cache. 
Low-bit quantization and model optimization reduce these demands~\cite{zandieh2026turboquant,liu2024mobilellm,lin2025awq,nielsen2024bitnet}, but substantial latency and energy are still spent moving weights, activations, and intermediate results between memory and compute~\cite{wolters2024memory}. 
Consequently, efficient edge inference depends not only on computational throughput, but also on how efficiently data can be moved, and where computation is performed.
\end{sloppypar}

\begin{figure}[tbp]
    \centering
    \includegraphics[width=\columnwidth]{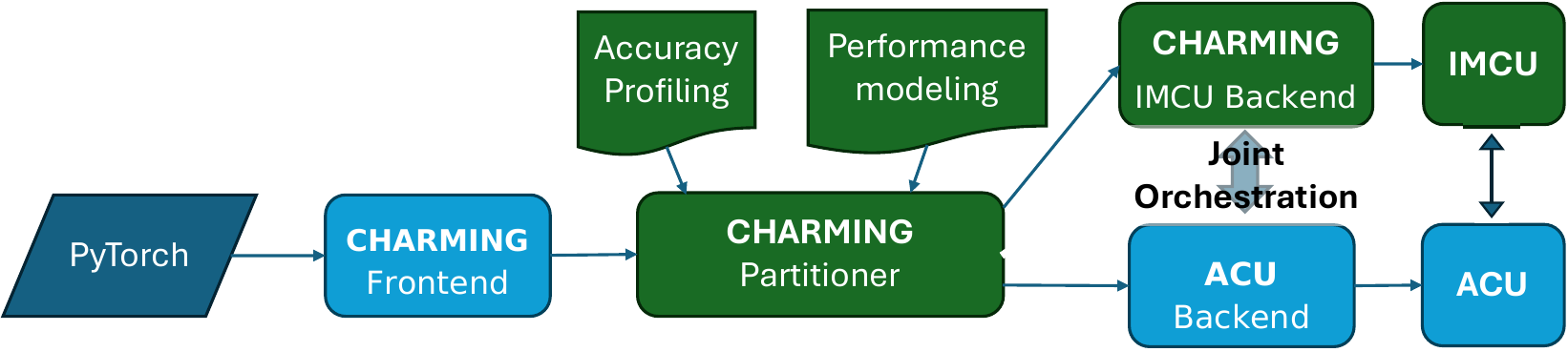}
    \vspace{-.25in}
    \caption{CHARMING design flow overview.}
    \label{fig:CHARMING-Overview}
    \vspace{-.25in}
\end{figure}
\looseness=-1
Heterogeneous reconfigurable platforms, such as AMD Versal ACAP, combine arrays of AI Engine (AIE) tiles with programmable logic and embedded processors, providing both high-throughput matrix computation and configurable data movement~\cite{dong2024eq,zhuang2023charm,zhuang2025aries,zhuang2024charm,dac23ACAP}. 
Prior work, CHARM, demonstrated that optimizing this data movement is critical to achieving high performance on AIE-based accelerators. 
By coordinating computation and communication across the programmable logic and AIE array, CHARM overlaps operand movement with matrix computation to reduce its impact on execution time~\cite{zhuang2023charm,zhuang2024charm}. 
Reconfigurable platforms provide a second opportunity: FPGA-based memory controllers can expose fine-grained control over DRAM commands and timing, enabling computation directly within the memory system. Prior work has demonstrated bulk-bitwise and arithmetic operations using commodity or minimally modified DRAM~\cite{oliveira2024mimdram,hajinazar2021simdram,seshadri2017ambit}, while recent in-DRAM designs extend these mechanisms to matrix computation~\cite{de2026count2multiply}. 

In this paper, we present CHARMING (CHARM with In-memory computiNG), a heterogeneous edge-AI architecture that combines AIE-based tensor processing with in-memory computation to reduce data movement and improve inference performance. As illustrated in Figure~\ref{fig:CHARMING-Overview}, CHARMING integrates the CHARM~\cite{zhuang2023charm,zhuang2024charm} compilation framework with a new in-memory computing backend to coordinate an AIE Compute Unit (ACU) and an In-Memory Compute Unit (IMCU) based on state-of-the-art DRAM PIM~\cite{de2026count2multiply}. Starting from a PyTorch model, the CHARMING frontend feeds a CHARMING partitioner that distributes transformer inference workloads across the two compute units. Offline accuracy profiling evaluates the effects of quantization and head selection on model accuracy, identifying computations suitable for reduced-precision in-memory execution. Subject to these accuracy constraints, analytical performance models determine the ACU--IMCU partition for both prefill and decode to maximize performance and improve energy efficiency through concurrent execution. The resulting partitions are passed to the existing CHARM(ACU) backend and the new CHARMING PIM backend. CHARMING jointly orchestrates these backends, integrating CHARM's existing coordination of programmable logic and host processing with the corresponding control, data movement, and synchronization required for in-memory computation. This unified execution flow enables the ACU and IMCU to operate concurrently while exploiting their complementary computational capabilities to improve performance and energy efficiency.

Our contributions are as follows:
\begin{itemize}
\item We propose CHARMING, a heterogeneous reconfigurable architecture that coordinates AIE-based tensor processing and in-DRAM computation to accelerate transformer inference through concurrent execution and reduced data movement.
\item We characterize the accuracy-performance tradeoffs of mixed-precision attention-head partitioning, demonstrating how selective quantization can preserve model accuracy while enabling heterogeneous execution.
\item We explore different partitioning strategies suitable for prefill and decode stages of these transformer workloads co-designed with the ACU and IMCU.
\item \begin{sloppypar} We demonstrate an end-to-end heterogeneous CHARMING validation prototype platform that integrates a Versal VCK5000 and an Alveo U200 for ACU and IMCU execution, respectively. \end{sloppypar}
\item We combine hardware measurements from the validation platform with detailed cycle-level simulation to evaluate CHARMING's performance and energy efficiency. CHARMING uses simulation only for in-memory execution as the extensions required for state-of-the-art in-DRAM computation are not available in commercial hardware.
\looseness=-1

\end{itemize}

CHARMING outperforms CHARM, as well as the GPU-based Jetson Xavier NX in latency: during prefill it achieves a $1.15$--$1.236\times$ speedup over CHARM and a $2.31$--$4.47\times$ speedup over the contemporary edge GPU, Jetson Xavier NX, and during decode it achieves a $1.06$--$1.44\times$ speedup over CHARM alone. To capture energy and latency together, we use the energy-delay product (EDP): in prefill, CHARMING reduces EDP by up to $24.7\%$ over CHARM and by up to $80.1\%$ over the Jetson Xavier NX.

\section{Prior Work}

\newsavebox{\motivA}\newsavebox{\motivB}\newlength{\motivH}
\begin{figure}[t]
    \centering
    \sbox{\motivA}{\includegraphics[height=1in]{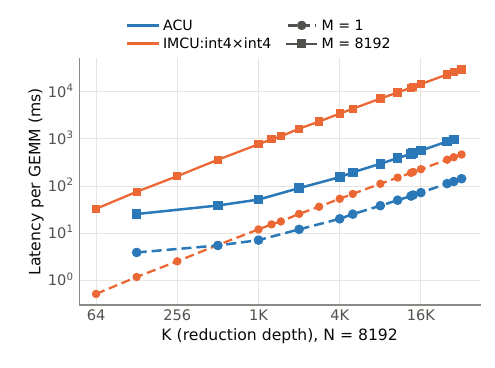}}%
    \sbox{\motivB}{\includegraphics[height=1in]{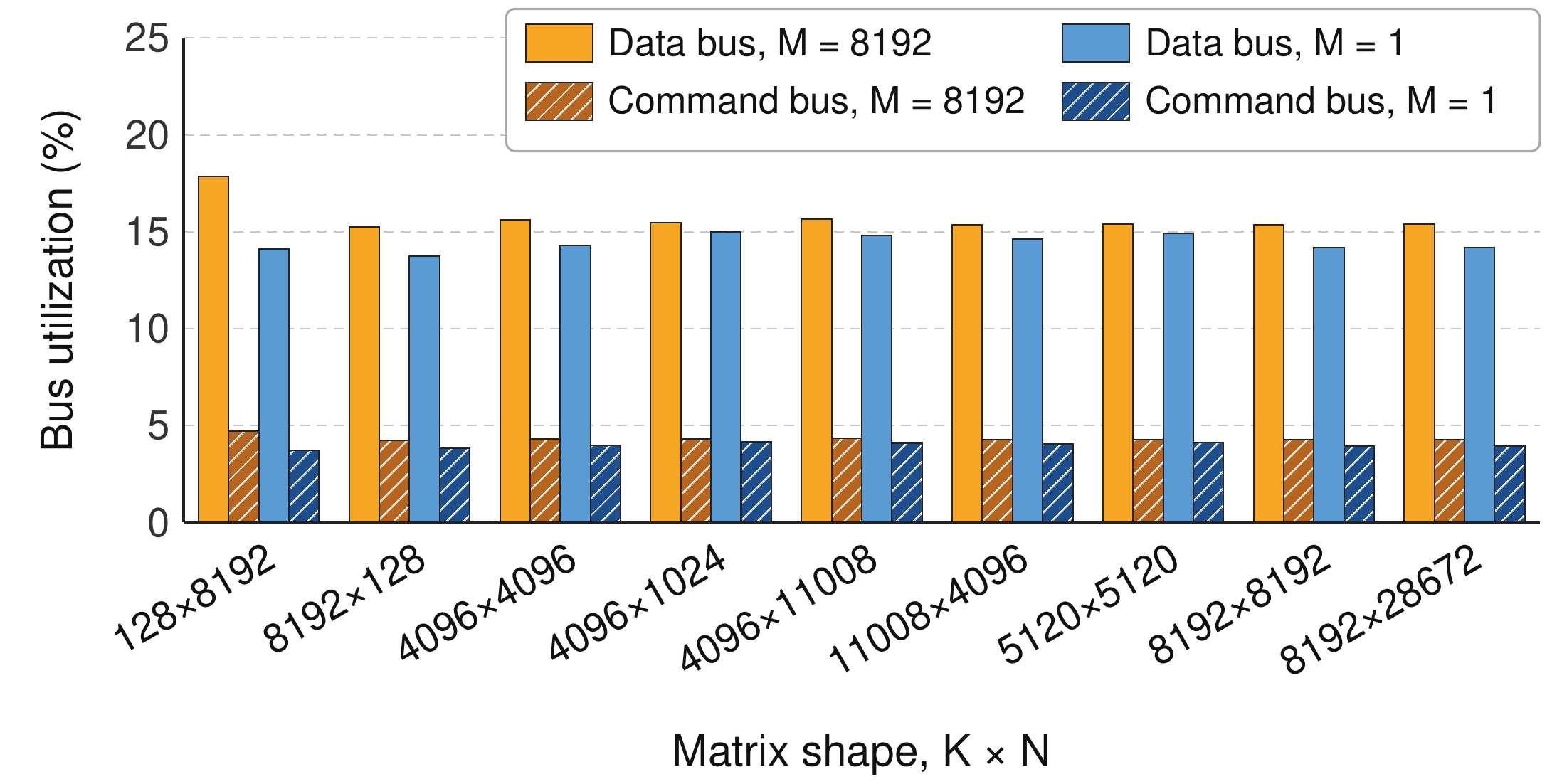}}%
    \setlength{\motivH}{\dimexpr(\linewidth-1.5em)*\number\dimexpr1in\relax
                        /\number\dimexpr\wd\motivA+\wd\motivB\relax\relax}%
    \sbox{\motivA}{\includegraphics[height=\motivH]{Figures/charmim_motivation.pdf}}%
    \sbox{\motivB}{\includegraphics[height=\motivH]{Figures/bus_utilization.pdf}}%
    \begin{subfigure}[t]{\wd\motivA}
        \centering
        \usebox{\motivA}
        \caption{Latency of the ACU and IMCU across reduction depth.}
        \vspace{-.1in}
        \label{fig:charmim-motivation}
    \end{subfigure}\hfill
    \begin{subfigure}[t]{\wd\motivB}
        \centering
        \usebox{\motivB}
        \caption{AIE-only compute unit's data and command-bus utilization for prefill ($M=8192$) and decode ($M=1$).}
            \vspace{-.1in}
        \label{fig:bus-utilization}
    \end{subfigure}
    \caption{Motivation for heterogeneous execution in CHARMING:
    (a) the IMCU favors shallow reductions, whereas the ACU favors deep ones;
    (b) the evaluated ACU workloads leave substantial data- and command-bus headroom.}
    \vspace{-.2in}
    \label{fig:motivation}
\end{figure}

The AMD/Xilinx Versal architecture integrates programmable logic with
an array of VLIW-SIMD AI Engine tiles, providing a reconfigurable
substrate for high-throughput matrix computation. Prior AIE-based
systems, including H-GCN, CHARM, and ARIES, optimize dataflow, tiling,
resource allocation, and workload scheduling for graph and dense matrix
operations; however, their operands must still be transferred between
external memory, programmable logic, and the AIE
array~\cite{zhang2022hgcn,zhuang2023charm,zhuang2024charm,zhuang2025aries}.

In parallel, Ambit introduced bulk-bitwise computation using DRAM row
operations, while SIMDRAM and MIMDRAM extended this foundation with
programmable arithmetic and greater subarray-level
parallelism~\cite{seshadri2017ambit,hajinazar2021simdram,
oliveira2024mimdram}. DRAM Bender provides an FPGA-based infrastructure
and command-level DRAM control for experimentally evaluating such
mechanisms, while Count2Multiply builds on bulk-bitwise operations to perform
matrix multiplication through massively parallel high-radix
counting~\cite{olgun2023dram,de2026count2multiply}.

Prior studies have also demonstrated heterogeneous co-execution between
conventional accelerators and PIM. NeuPIMs targets
high-throughput batched LLM serving and enables concurrent NPU--PIM
execution through dual row buffers and sub-batch interleaving, requiring
independent requests and specialized memory
modifications~\cite{heo2024neupims}. Similarly, IANUS targets datacenter
LLM inference using a unified NPU--PIM memory system and dedicated
workload-mapping and memory-access
scheduling~\cite{seo2024ianus}. H4H considers an edge-oriented NPU--CIM
platform, but applies neural architecture search to distribute layers of
hybrid CNN--Transformer models rather than concurrently partitioning
attention within an LLM layer~\cite{zhao2025h4h}. In contrast, CHARMING
enables fine-grained, head-level co-execution within a single inference
request, allowing the ACU and IMCU to process the same attention stage concurrently while reducing
external data movement.

Head-wise splitting is the standard intra-layer parallelism for Transformers:
Megatron-LM gives each GPU a subset of heads, since heads are independent until
the output projection~\cite{shoeybi2019megatron,narayanan2021megatron,pope2023efficiently}.
HexGen and Hetis extend it to uneven splits across heterogeneous
GPUs~\cite{jiang2024hexgen,mo2025hetis}.
All of them run the same operation in the same number format on every device.
CHARMING splits by head across two different substrates, FP32 tensor
processing and in-DRAM integer computation, whose per-head costs differ by
stage and precision.
\vspace{-.1in}

\section{Motivation}
\label{sec:motivation}

The ACU and IMCU exhibit complementary performance characteristics. As shown in
Fig.~\ref{fig:charmim-motivation}, the IMCU is more efficient for
operations with small reduction workloads, for instance \(M = 1\) and
\(K < 512\), where the overhead of transferring operands to the AIE array
dominates execution: at \(M = 1\), \(K = 128\), the IMCU is \(3.3\times\)
faster than the ACU. As \(M \times K\) increases, the ACU's higher arithmetic
throughput amortizes this overhead and eventually outperforms the IMCU; the
two cross near \(K = 512\) at \(M = 1\), and at \(M = 8192\) the ACU is faster
at every measured depth. Consequently, assigning every operation to either
unit alone leaves part of the available heterogeneous compute capability
underutilized.

Figure~\ref{fig:bus-utilization} further shows that the evaluated ACU
workloads use only \(13.7\)--\(17.9\%\) of the DRAM data bus and
\(3.5\)--\(4.7\%\) of the command bus. This complementary resource behavior
motivates CHARMING: compute-intensive operations remain on the ACU,
whereas suitable memory-bound operations execute in place on the IMCU.
The two units can therefore operate concurrently, reducing external data
movement and exploiting otherwise unused memory-system capacity.

\section{System Design}
This section presents the overall CHARMING architecture, parallelization, and memory-mapping strategies, and describes how the system orchestrates data movement across its components.
\subsection{CHARMING System Architecture}
\begin{figure}
    \centering
    \includegraphics[width=.8\linewidth]{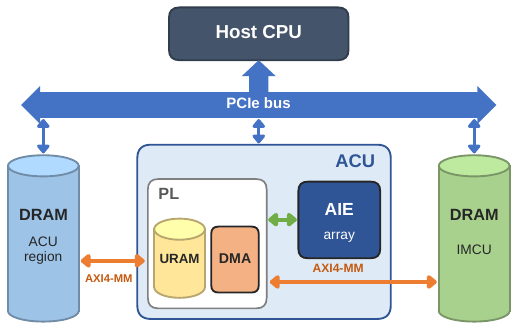}
    \vspace{-.1in}
    \caption{CHARMING Architecture Overview.}
    \vspace{-.2in}
    \label{fig:architecture}
\end{figure}

Figure~\ref{fig:architecture} shows CHARMING, which couples an AIE compute
unit (ACU) with an in-memory compute unit (IMCU) so that both work on the
same model concurrently. Each unit works on its own DRAM region, and the host CPU
accesses both regions over PCIe.  Section~\ref{sec:pipeline}
describes how the workload is distributed between them.

The host loads the bitstream once at initialization. At run time it only
supplies data and commands: it writes the inputs to DRAM, sets which heads
each unit computes and where their data reside, and starts the ACU and the
IMCU. Inside the ACU, the DMA in the PL reads operand tiles from the ACU's
DRAM over AXI4-MM, buffers them on chip, and streams them to the AIE array
over AXI4-Stream. Results stream back to URAM in the PL for accumulation.
The DMA then writes each finished tile in a single pass: the heads assigned
to the IMCU are quantized from FP32 to N-bit and written to the IMCU's DRAM,
while the remaining heads stay FP32 and return to the ACU's DRAM. DRAM
transfers use AXI4-MM because they need addressed accesses; the AIE links
use AXI4-Stream because they do not. The IMCU then computes on its heads in
place with row-level DRAM commands.

Because the IMCU computes in place, it works mainly through the DRAM command
bus, which the ACU leaves mostly idle (Section~\ref{sec:motivation}).
The two units therefore run in parallel with little contention for
data-bus bandwidth. 
No single device, however, currently combines AIE arrays with in-DRAM
compute, and commodity DRAM does not support the in-subarray inversion that in-memory
multiplication requires. We therefore prototype CHARMING on two boards, as
described in Section~\ref{sec:experimental setup}.

\subsection{CHARMING Dataflow}

\begin{figure}
    \centering
    \includegraphics[width=1\linewidth]{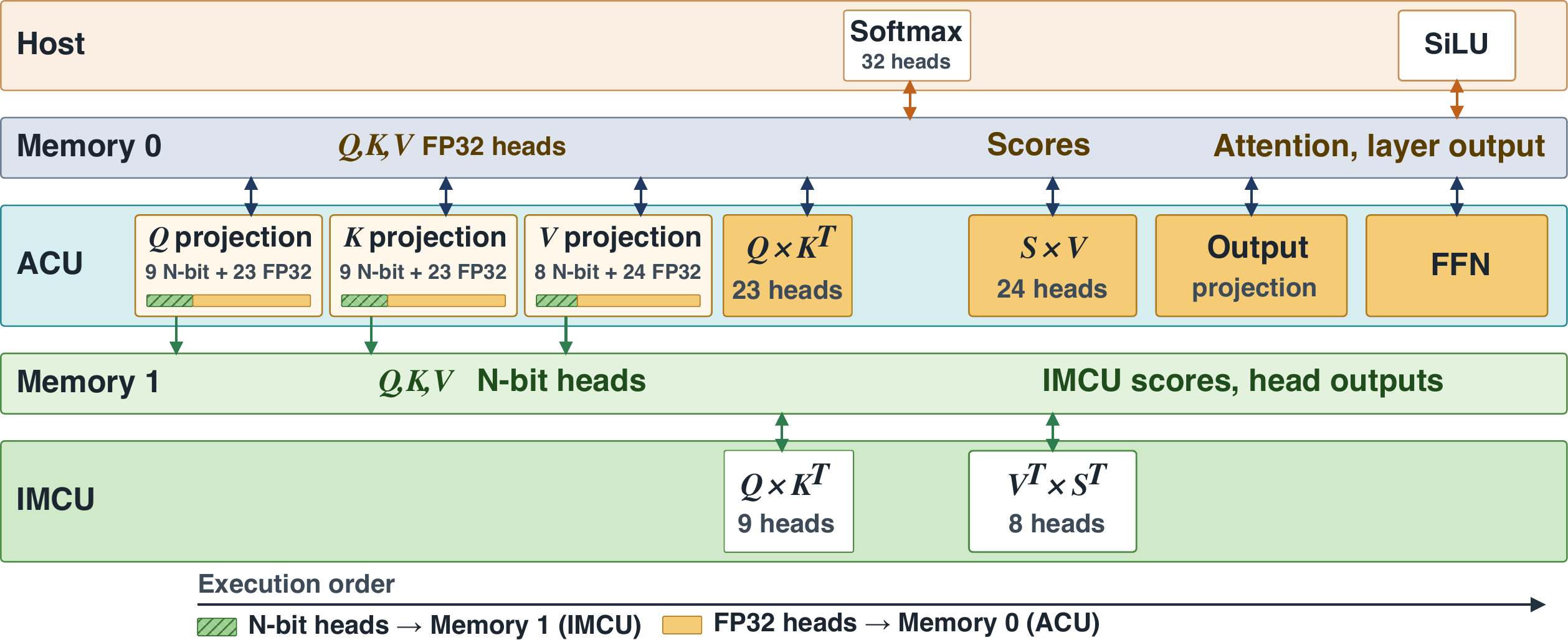}
    \vspace{-.25in}
    \caption{CHARMING dataflow (left to right) for an example layer (LLaMA 2 7B) during prefill. Data is partitioned during projection on the ACU, with FP32 data supporting ACU execution stored in memory 0 and in-memory compute quantized data stored in memory 1.}
    \label{fig:dataflow}
    \vspace{-.1in}
\end{figure}

Figure~\ref{fig:dataflow} demonstrates the flow of a LLaMA 2 7B decoder layer, with 32-head multi-head attention (MHA),
through the CHARMING architecture (Figure~\ref{fig:architecture}) in execution order, across the host, the ACU, the IMCU, and
two memory regions. 
The ACU computes each $Q$, $K$, and $V$ projection in a single FP32 GEMM pass.
CHARMING writes the heads assigned to the IMCU in quantized format (precision int$_N$) to
Memory~1 and the remaining heads in FP32 to Memory~0, so quantization needs
neither an extra pass nor a bitstream reconfiguration. The output projection
and FFN run on the ACU; softmax and SiLU run on the host.

$Q\times K^\top$ and $S\times V$ are split head-wise because heads are the
natural unit of independence in attention. Because the two units run
concurrently, each stage costs the longer of its two paths, and the head-split
optimizer balances them. For $Q\times K^\top$, the IMCU receives
\begin{equation}
N_{opt}^{QK} = \max\!\left(0,\ \left\lfloor
  \frac{T_{v} + (H+1)\, t_{qk}^{\mathrm{ACU}}}
       {t_{qk}^{\mathrm{ACU}} + t_{qk}^{\mathrm{IMCU}}}
\right\rfloor\right)
\label{eq:nopt_qk}
\end{equation}
heads, where $T_v$ is the $V$ projection time, which overlaps this stage on
the ACU, $H$ is the number of heads, and $t_{qk}^{\mathrm{ACU}}$ and
$t_{qk}^{\mathrm{IMCU}}$ are the per-head latencies estimated by Equations~\ref{eq:acu-latency} and~\ref{equ:latency} respectively.
The optimal heads in IMCU for $S\times V$ are
\begin{equation}
N_{opt}^{SV} = \max\!\left(0,\ \left\lfloor
  \frac{(H+1)\, t_{sv}^{\mathrm{ACU}}}
       {t_{sv}^{\mathrm{ACU}} + t_{sv}^{\mathrm{IMCU}}}
\right\rfloor\right)
\label{eq:nopt_sv}
\end{equation}
For LLaMA 2 7B with 4-bit quantization of both matrices (int4$\times$int4), the partitioning produces 9 and 8 IMCU heads,
from Eqs.~\ref{eq:nopt_qk} and \ref{eq:nopt_sv}, respectively.
The IMCU, however, computes $S \times V$ in the form
$S\times V = (V^{\top}S^{\top})^{\top}$ to exploit the operand asymmetry of
Count2Multiply~\cite{de2026count2multiply}. Count2Multiply (C2M) stores one operand in DRAM rows and broadcasts the
elements of the other across the in-memory compute array, performing
multiplication and accumulation through parallel counting, with one
matrix--vector operation per row of the broadcast operand. Storing $S^{\top}$
and broadcasting the smaller $V^{\top}$ therefore reduces these operations
from $n$ to $d_h$, where $n$ is the sequence length and
$d_h$ is the head dimension, and raises DRAM row utilization from
$d_h/n$ to 100\%. This is what makes offloading $S \times V$
viable in prefill. The softmax writes $S$ in column-major order, so each row
of $S^{\top}$ is contiguous, and the IMCU's output rows form $SV$ in
column-major order, the layout the ACU reads, so no transpose is needed
afterwards.



\subsection{Pipelining and Parallelization of Transformer Layers}
\label{sec:pipeline}

The MHA decomposes into $H$ independent per-head matrices, so any head can be placed on either unit without affecting the others (Figure~\ref{fig:dataflow}). Each $Q$, $K$, and $V$ projection is a
single pass whose integrated kernel writes the IMCU heads in N-bit form, so the IMCU starts its $Q\times K^\top$ heads as soon as the $Q$ and $K$ passes complete, while the ACU computes the $V$ projection and its own
$Q\times K^\top$ heads. During this stage the ACU streams its operands over the data bus, whereas the IMCU computes in place (Figure~\ref{fig:dataflow}). $S\times V$ follows the same pattern, costing $\max\big((H-N_{opt}^{SV})\,t_{sv}^{\mathrm{ACU}},\; N_{opt}^{SV}\,t_{sv}^{\mathrm{IMCU}}\big)$.

Like $Q,K,V$ projection, the output projection and FFN run on the ACU alone as well, as the IMCU is competitive only when the product of output rows M and reduction depth K is small.
IMCU latency grows as outlined in Equation~\ref{equ:latency}:
\begin{equation}
t^{\mathrm{IMCU}}(M,K,N)
  = \frac{M\,\kappa(A,B)\,K\,\varphi(K)}{W_{\mathrm{tot}}}
    \left\lceil \frac{N}{N_{\mathrm{row}}} \right\rceil ,
\label{equ:latency}
\end{equation}
where $M$ is the number of output rows, $K$ the reduction depth,
$N$ the output width, $N_{\mathrm{row}} = 8192$ the output columns in one DRAM
row, and $W_{\mathrm{tot}}$ the banks available to the IMCU (8 per
pseudo-channel; 128 with all 16, 112 or 96 with 14 or 12). The ACU latency
follows the CHARM model~\cite{zhuang2023charm}:
\begin{equation}
T_{\mathrm{comp}}
= \frac{X \times Y \times Z \times T_I \times T_K \times T_J}
       {\mathrm{MAC}\ \times \mathrm{Eff}},
\label{eq:acu-compute}
\end{equation}
\begin{equation}
t^{\mathrm{ACU}}
= \max\!\left(T_L, T_R, T_{\mathrm{comp}}\right)
  \times T_X \times T_Y \times T_Z ,
\label{eq:acu-latency}
\end{equation}
where $X, Y, Z$ are the on-chip loop counts, $T_I, T_K, T_J$ the tile
dimensions per AIE, $\mathrm{MAC}$ the MACs per AIE per cycle, $\mathrm{Eff}$
the achieved pipeline efficiency, $T_L$ and $T_R$ the operand transfer
latencies, and $T_X, T_Y, T_Z$ the outer tiling counts. 

\subsubsection{Precision Cost}
Count2Multiply~\cite{de2026count2multiply} computes a product of an $A$-bit
operand and a $B$-bit operand in DRAM. The $B$-bit operand is bit-sliced into $B$ binary planes held in DRAM rows, each acting as a per-column predicate that advances a Johnson counter by the value of the $A$-bit operand. The single-bank cost per unit of $MK$ is therefore
\begin{equation}
\kappa(A,B) \approx c\,A\,(B-1)^{3/2},
            \qquad c = 6.59\times10^{-4}\ \mathrm{ms}.
\label{equ:kappa}
\end{equation}

\subsubsection{Amortization Factor}
Carries between counter digits are deferred and settled in bulk, a largely
fixed cost per counter that a long accumulation amortizes over more inputs:
\begin{equation}
\varphi(K) = 1 - \frac{174}{K + 390}
\label{equ:phi}
\end{equation}
%
We fit the constants of Equations~\ref{equ:kappa} and~\ref{equ:phi} to
simulated per-head latencies for $A\in\{4,5\}$, $B\in\{4,5,6\}$ and
$K\in[64, 32768]$. Equations~\ref{equ:latency}--\ref{equ:phi} then lie within 5.1\% of every simulated latency.

\subsection{Head-Split Mapping}
\label{sec:mapping}
 
Equations~\ref{eq:nopt_qk}--\ref{eq:nopt_sv} treat the IMCU time as linear in the head count. The IMCU, however, issues row operations to $W_{\mathrm{tot}}$ banks in parallel, so its time is a step function of the head count.
Grouped-Query Attention (GQA) adds a second consideration: the $H$ query heads share $H_{kv}$ key--value heads in groups of $g = H/H_{kv}$, so a group whose query heads run on both units needs its key--value head on both, in $N$ bits for the IMCU and in FP32 for the ACU.
We therefore generalize Equations~\ref{eq:nopt_qk}--\ref{eq:nopt_sv} into an exact cost model, valid for MHA ($g = 1$) and GQA, and search it exhaustively offline per model, phase (prefill/decode), and
precision with Algorithm~\ref{alg:mapping}.

\paragraph{Cost model.}
Let $N^{QK}$ and $N^{SV}$ be the numbers of heads whose $Q\times K^{\top}$ and $S\times V$ run on the IMCU. The two stages are split independently, since host softmax converts a head's scores to $N$-bit formats. 
The ACU computes $Q$, $K$ and $V$ in one pass each. The $Q$ pass outputs $H d_h$ columns and takes a constant time $T_q$.
The $K$ and $V$ passes each output $d_h H_{kv}$ columns, padded to tiles of $N_{\mathrm{acc}} = 1024$ columns, and take
\begin{equation}
T_k = T_v = \tau_{kv} \left\lceil d_h H_{kv} / N_{\mathrm{acc}} \right\rceil ,
\label{eq:kpass}
\end{equation}
where $\tau_{kv}$ is the time of one tile column ($t^{\mathrm{ACU}}$ of Equation~\ref{eq:acu-latency} for an $n \times d \times N_{\mathrm{acc}}$ GEMM, with $d$ the model dimension).
Each key--value head is computed once, in FP32, regardless of where its query heads run: the memory mapper quantizes a head's output columns only if the IMCU reads them and writes them in both formats if both units do (Algorithm~\ref{alg:mapping}).
The projection times therefore do not depend on the split, and GQA adds no ACU cost.
 
Equation~\ref{equ:latency} spreads the $R = M\lceil N/N_{\mathrm{row}}\rceil$
row operations of one head over the banks, at $u = \kappa(A,B)\,K\,\varphi(K)$
each. Counting whole rounds of bank operations,
\begin{equation}
t^{\mathrm{IMCU}}_{qk} = u_{qk} \left\lceil R_{qk} N^{QK} / W_{\mathrm{tot}} \right\rceil ,
\label{eq:banks}
\end{equation}
and likewise for $S\times V$. In prefill this is within one row operation of
Equation~\ref{equ:latency}; in decode, where $R_{qk} = 1$, up to
$W_{\mathrm{tot}}$ $Q\times K^{\top}$ heads cost a single row operation. The
shape $(M,K,N)$ is $(n, d_h, n)$ for $Q\times K^{\top}$ and, transposed,
$(d_h, n, n)$ for $S\times V$; in decode it is $(1, d_h, n)$ and $(d_h, n, 1)$ respectively.
 
The ACU and IMCU synchronize after each attention stage
(Figure~\ref{fig:dataflow}): $S\times V$ starts on either unit only once all
heads have finished $Q\times K^{\top}$ and softmax, and the output projection
only once all heads have finished $S\times V$. Within the first stage, the IMCU
starts $Q\times K^{\top}$ when the $Q$ and $K$ passes end, at
$t_1 = T_q + T_k$, while the ACU first computes the $V$ pass.
The layer takes
\begin{multline}
T = t_1 + \max\!\big(T_v + (H - N^{QK})\,t^{\mathrm{ACU}}_{qk},\; N^{QK}\times t^{\mathrm{IMCU}}_{qk}\big) \\
    + \max\!\big((H - N^{SV})\,t^{\mathrm{ACU}}_{sv},\; N^{SV}\times t^{\mathrm{IMCU}}_{sv}\big) + T_{\mathrm{post}},
\label{eq:layer}
\end{multline}
where $T_{\mathrm{post}}$ is the output projection and FFN on the ACU. 
 
\paragraph{Search.}
$T$ depends only on $(N^{QK}, N^{SV})$, not on which heads are offloaded.
We evaluate Equation~\ref{eq:layer} for all $(H+1)^2$
pairs, at most $4{,}225$ for the 64-head models, and keep the fastest,
preferring fewer offloaded heads on ties. All decoder layers of a model have
the same shapes, so one search serves every layer.
With $R_{qk}N^{QK}$ and $R_{sv}N^{SV}$ multiples of $W_{\mathrm{tot}}$,
Equation~\ref{eq:layer} separates into two problems
$\min_N \max\!\big(\alpha + (H - N)\,t^{\mathrm{ACU}},\, N t^{\mathrm{IMCU}}\big)$,
whose integer minimizers are Equations~\ref{eq:nopt_qk}--\ref{eq:nopt_sv}, with
$\alpha = T_v$ and $\alpha = 0$.

\paragraph{Head selection.}
Within each layer, we rank the query heads by their \emph{damage score} (Section~\ref{sec:head-selection}) at precision $(A,B)$ and offload the $N^{QK}$ and $N^{SV}$ best-ranked heads.
Under GQA, heads are selected individually rather than by group; since the latency depends only on the head counts, this choice affects accuracy alone.
Section~\ref{sec:head-selection} compares the accuracy of this selection with that of randomly chosen heads.
 
\begin{algorithm}[t]
\caption{Offline head-split mapping}
\label{alg:mapping}
\begin{algorithmic}[1]
\Require model $(H, H_{kv}, d_h, \dots)$, phase, length $n$, precision $(A,B)$,
         per-layer damage-score ranking of the query heads
\Ensure per layer, IMCU heads $\mathcal{S}^{QK}, \mathcal{S}^{SV}$; per pass $P$, the
        head-index sets $\mathcal{I}_P$ (IMCU) and $\mathcal{F}_P$ (ACU)
\State $T_q, \tau_{kv}, t^{\mathrm{ACU}}_{qk}, t^{\mathrm{ACU}}_{sv}, T_{\mathrm{post}} \gets$
       ACU model (Eqs.~\ref{eq:acu-compute}--\ref{eq:acu-latency})
\State $u_{qk}, R_{qk}, u_{sv}, R_{sv} \gets$ IMCU model (Eqs.~\ref{equ:latency}--\ref{equ:phi})
\State $(N^{QK}, N^{SV}) \gets \arg\min_{0 \le N^{QK}, N^{SV} \le H} T$
       \Comment{Eq.~\ref{eq:layer}}
\For{each decoder layer}
  \State $\mathcal{S}^{QK}, \mathcal{S}^{SV} \gets$ the layer's $N^{QK}$ and $N^{SV}$ best-ranked heads
  \For{$(P, \mathcal{S}) \in \{(Q, \mathcal{S}^{QK}), (K, \mathcal{S}^{QK}), (V, \mathcal{S}^{SV})\}$}
    \State $\mathcal{I}_P \gets \mathcal{S}$ if $P = Q$, else
           $\{j : \mathcal{G}_j \cap \mathcal{S} \neq \emptyset\}$
    \State $\mathcal{F}_P \gets$ heads not in $\mathcal{S}$ if $P = Q$, else
           $\{j : \mathcal{G}_j \not\subseteq \mathcal{S}\}$
  \EndFor
\EndFor
\end{algorithmic}
\end{algorithm}

Lines~6--8 compute, for each pass $P$, the heads read by the IMCU ($\mathcal{I}_P$) and by the ACU ($\mathcal{F}_P$), where $\mathcal{G}_j$ denotes the query heads sharing key--value head $j$.
Before each layer, the host writes these index sets to the memory mapper's AXI4-Lite control registers (Figures~\ref{fig:dataflow} and~\ref{fig:compilation-flow}).
Each head occupies $d_h$ consecutive output columns, so the mapper identifies the head of every column from its index.
As each output tile leaves the ACU, the mapper quantizes the columns of the heads in $\mathcal{I}_P$ and writes them to the IMCU in $N$ bits, and writes the columns of the heads in $\mathcal{F}_P$ to the ACU in FP32.
A split group's key--value head is in both sets and is written in both formats from a single GEMM result.
Neither the weights nor the projection change with the selection; a different selection only rewrites the index sets.

\section{Experimental Setup}
\label{sec:experimental setup}
CHARMING is evaluated in a two-step process. We use a full hardware emulation approach to validate our designs and we use a mixture of hardware execution and simulation for performance and energy characterization~\cite{olgun2023dram,zhuang2023charm}. We also evaluate the effect of quantization. We use GSM8K~\cite{cobbe2021gsm8k}, a dataset of high-quality, linguistically diverse grade school math word problems for benchmarking. Three LLMs with different model structures are evaluated in the experiment: Llama-3.1-8B-Instruct~\cite{grattafiori2024Llama3}, Llama-2-13B-chat~\cite{touvron2023Llama2} and Mixtral-8x7B-Instruct-v0.1~\cite{jiang2024mixtral}. We use NVIDIA H200, PyTorch 2.14 and CUDA 13 for all quantization evaluation experiments.

\begin{table}[t]
\centering
\caption{Memory organization and architectural parameters of the prototype
hardware and of the target system.}
\label{tab:platform}

\footnotesize
\setlength{\tabcolsep}{3pt}
\renewcommand{\arraystretch}{1.10}

\begin{tabularx}{\columnwidth}{@{}
    >{\centering\arraybackslash}p{0.23\columnwidth}
    |
    >{\centering\arraybackslash}X
    @{}}
\toprule
\multicolumn{2}{@{}c@{}}{\textbf{Prototype hardware}} \\
\midrule

AIE compute unit &
VCK5000 Versal Development Card; VC1902 AI Core, 7\,nm;
400 AIE tiles at 1.25\,GHz; PL kernel at 230\,MHz;
16\,GB DDR4-3200 (4 channels $\times$ 4\,GB)
\\
\midrule

In-memory compute unit &
Alveo U200 (XCU200, Virtex UltraScale+, 16\,nm);
DRAM-Bender controller; 64\,B write register;
PCIe Gen3\,$\times$8 link
\\
\midrule

DDR4 (U200) &
SK hynix HMA82GR7AFR8N-UH, 16\,GB 2Rx8 RDIMM (8\,Gb x8 devices);
rated 2400\,MT/s, operated at 1333\,MT/s ($t_{CK}=1.5$\,ns); one rank used;
4 BG $\times$ 4 banks = 16 banks; 64-bit bus;
8\,kB row per rank (1\,kB per device); 
$t_{RCD}=t_{RP}=10$, $t_{RAS}=22$, $t_{FAW}=9$, $t_{RRD,S}/t_{RRD,L}=3/4$,
$t_{RFC}=234$, $t_{REFI}=5200$ (in $t_{CK}$)
\\

\midrule

Host CPU &
Intel Xeon w3-2423; 6C/12T at up to 4.2\,GHz;
32\,GB DDR5; Ubuntu 22.04; Linux 5.15
\\
\midrule[\heavyrulewidth]
\multicolumn{2}{@{}c@{}}{\textbf{Target system}} \\
\midrule

HBM2 &
\texttt{HBM2\_4Gb\_1x64}; 16 pseudo-channels $\times$ 8 banks
(4 BG $\times$ 2) = 128 banks, 4\,GB; 64-bit bus per pseudo-channel;
1\,kB row; 2000\,MT/s;
$t_{RCD}=t_{RP}=t_{CL}=7$\,ns, $t_{RAS}=17$\,ns, $t_{FAW}=15$\,ns,
$t_{RRD,S}/t_{RRD,L}=2/3$\,ns, $t_{RFC}=260$\,ns, $t_{REFI}=3.9\,\mu\mathrm{s}$
\\

\midrule

On-chip Arm CPU &
VCK190 (VC1902); dual-core Arm Cortex-A72 at 1.35\,GHz; 8\,GB DDR4;
PetaLinux 2021.1; Linux 5.10
\\
\bottomrule
\end{tabularx}
\end{table}

\begin{figure}[t]
  \centering
  \begin{subfigure}[t]{0.64\linewidth}
    \centering
    \includegraphics[width=\linewidth]{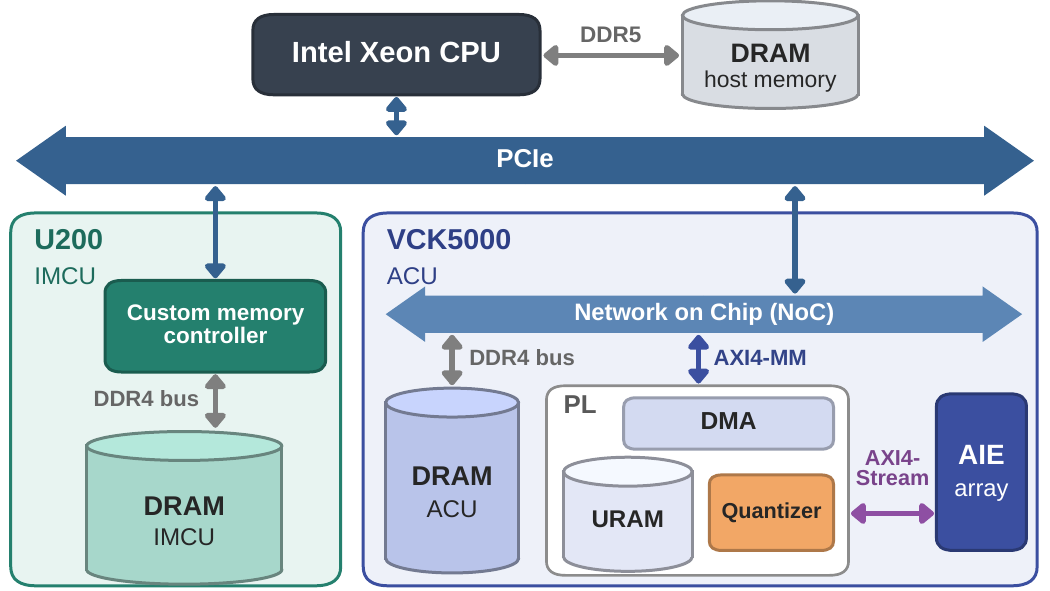}
    \caption{Hardware setup.}
    \label{fig:setup}
  \end{subfigure}\hfill
  \begin{subfigure}[t]{0.34\linewidth}
    \centering
    \adjustimage{trim={0.28\width} 0 {0.02\width} 0, clip, height=1.2in}{Figures/setup_photo}
    \caption{Platform}
    \label{fig:setup-hw}
  \end{subfigure}
  \vspace{-.1in}
  \caption{CHARMING hardware: (a) block diagram, (b) VCK5000 (ACU),
           Alveo U200 (IMCU) and Intel Xeon host.}
    \vspace{-.1in}
  \label{fig:setup-all}
\end{figure}

\subsection{Hardware Validation Flow}

CHARMING is evaluated using a heterogeneous hardware validation platform comprising two FPGA cards attached to a common x86 host over PCIe, as shown in Fig.~\ref{fig:setup-all}. The accelerators provide complementary execution substrates: an AIE-based spatial dataflow array and a DRAM-based compute-in-memory unit. In the primary configuration, data exchange between the accelerators is staged through shared host memory over PCIe. The host coordinates execution and performs operations not implemented natively by either accelerator. The major components in our hardware platform are summarized in Table~\ref{tab:platform}.

\subsubsection{AIE Compute Unit}

The AIE compute unit is implemented on an AMD Versal AI Core VC1902 adaptive SoC, using the VCK5000 accelerator
platform~\cite{zhuang2023charm}. The VC1902 integrates an AI Engine (AIE)
array that provides the dense GEMM capability used by CHARMING.
A PL kernel interfaces the AIE array with on-card DDR4: it tiles the operand
matrices, quantizes the operands assigned to the in-memory compute
path and
writes results to DDR4 through memory-mapped AXI4. A separate AXI4-Lite
interface provides configuration and control from the host. 

\subsubsection{In-Memory Compute Unit}
The in-memory compute unit is implemented on an AMD Alveo U200 (XCU200) accelerator
using a soft DDR4 memory controller implemented in programmable
logic~\cite{olgun2023dram}.
 Each DRAM bit column holds one modulo-10 Johnson counter. Each counter occupies five DRAM rows, with additional rows used for carry detection and scratch space.

\subsubsection{Host}
The host provides system memory for staging data and performs operations not
currently implemented by either accelerator. These include floating-point to
integer quantization, integer to floating-point de-quantization, and softmax. We evaluated the VCK190's embedded Arm processor and the Intel Xeon processor (Table~\ref{tab:platform}) for non-GEMM operations such as softmax, as shown in Figure~\ref{fig:softmax}.

\subsection{Emulation Approach}
Two aspects of the proposed CHARMING architecture are emulated by the present evaluation platform. First, while charge-sharing operations can be realized in commodity DDR4 using an FPGA-based soft memory controller, as implemented on the U200 (Table~\ref{tab:platform}), newer FPGA platforms increasingly rely on hardened memory controllers to support the higher speeds and bandwidths of contemporary memory technologies. These controllers do not expose the fine-grained command and timing control required for in-DRAM computation. Consequently, no currently available FPGA platform combines the necessary command-level DRAM control with an AIE array. We therefore implement the AIE and in-memory compute units on separate FPGA platforms, coupled through shared host memory. This configuration allows us to validate concurrent heterogeneous execution, coordination, and data exchange while emulating the tightly coupled, single-device integration envisioned by CHARMING.

Second, all required in-memory operations can be realized on the hardware platform, but some are not implemented using the most efficient mechanisms proposed or assumed by the proposed architecture~\cite{seshadri2017ambit,tegge2026low,LISA16}. In particular, inversion is functionally implemented in DRAM, but the present implementation requires additional data movement to make the complemented row available to subsequent operations. Directly incorporating this prototype-specific overhead would therefore substantially overstate the cost of inversion in an optimized implementation. For system-level evaluation, we therefore simulate the IMCU on a high-bandwidth
memory, HBM2 (Table~\ref{tab:platform}), using gem5-CIMD simulator~\cite{CIMD26}, while using the hardware realization to validate its underlying functionality. We describe these two aspects of the emulation methodology below.

\subsubsection{Separate devices} The consequence of separating the ACU and IMCU onto separate FPGA
devices is that operands exchanged between the two units must traverse
an external communication channel, PCIe. Due to the
software architecture of these platforms, data transfers are staged
through host memory before reaching the consuming device, introducing
additional communication and orchestration latency. In the proposed
integrated CHARMING architecture (Figure~\ref{fig:architecture}), the
ACU and IMCU retain independent memories but share a common
communication infrastructure and on-chip host processor for
orchestration, eliminating the additional host-mediated transfers
required by the prototype.

As illustrated in Figure~\ref{fig:setup-all}, the two-device evaluation platform allows
us to validate concurrent heterogeneous execution and data exchange.
The performance evaluation, however, represents the proposed
integrated architecture rather than the additional communication
overhead of the prototype. As demonstrated in Figure~\ref{fig:bus-utilization},
CHARM-style AIE execution utilizes only 14--18\% of
the DRAM data bus and 4--5\% of the command bus
for the evaluated workloads, leaving substantial
capacity available for in-memory operations.
These measurements support our assumption that ACU
and IMCU execution can overlap with limited interference
from contention for shared DRAM resources. Accordingly,
we characterize ACU execution and data movement using hardware
measurements and IMCU execution using cycle-level simulation,
combining their execution times under this concurrency assumption.
The reported performance excludes the additional PCIe and
host-memory staging overhead introduced by the separate FPGA
devices.

\subsubsection{Inversion}
Count2Multiply~\cite{de2026count2multiply} assumes Ambit-style
inversion~\cite{seshadri2017ambit}, which uses dual-contact cells
to access the complementary bitline of a sense amplifier.
Commodity DRAM lacks these cells but can perform inversion across
neighboring subarrays that share sense amplifiers~\cite{fcdram}.
However, the inverted result resides in the neighboring subarray,
and RowClone cannot return it to the original subarray because
it supports only intra-subarray copying.

To validate Count2Multiply on commodity DRAM, we perform inversion
in the neighboring subarray, at an address offset of 512 rows
on our module~\cite{fang2026memory,fang2026yavin}. The host then reads
the inverted row through the PCIe channel and
writes it back to the computing subarray.
This enables functional validation but introduces substantial
overhead, particularly because Count2Multiply requires inversion
of both mask and counter bits during counter updates. This
prototype-specific overhead is excluded from the simulated IMCU
performance, which assumes native Ambit-style inversion.



\subsection{CHARMING Compilation Flow}
\begin{figure}
    \centering
    \includegraphics[width=1\linewidth]{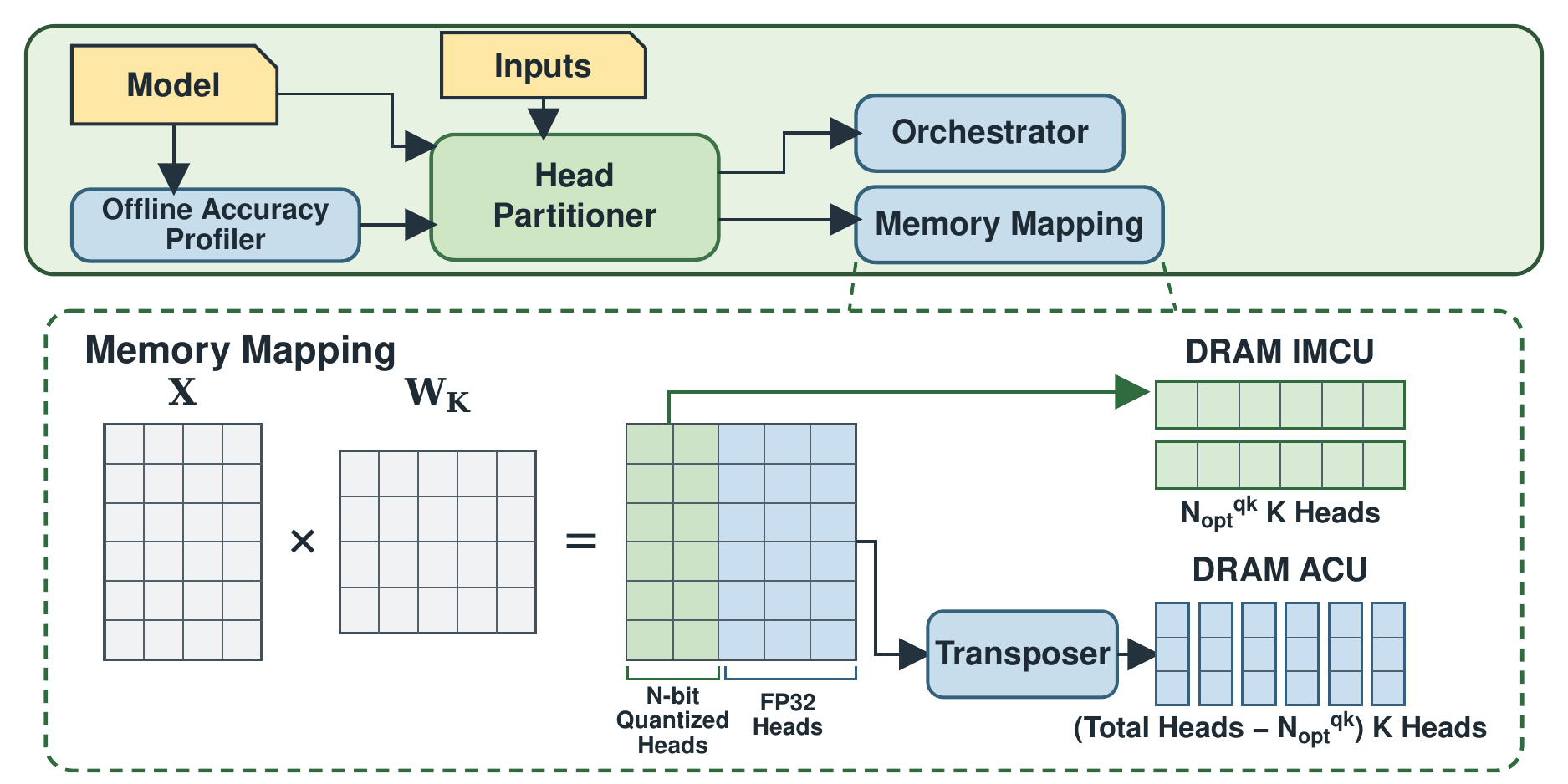}
    \vspace{-.3in}
    \caption{CHARMING Partitioner Detailed Flow with address mapping of $K~Projection = X \times W_{K}$ as an example. }
    \vspace{-.2in}
    \label{fig:compilation-flow}
\end{figure}
Figure~\ref{fig:compilation-flow} summarizes CHARMING’s system-level compilation flow. Given the model and input configuration, the host selects the operand precision, partitions the attention heads between the AIE compute unit (ACU) and in-memory compute unit (IMCU), generates the execution schedule, and maps the corresponding operands to memory. CHARMING uses AXI4 memory-mapped interconnects for data movement and AXI4-Lite registers for runtime control. Through these registers, the host configures the number of heads assigned to each compute unit, the required quantization mode, and the scheduling of subsequent operations.

For each head, the host determines whether the projection output is quantized to an \(N\)-bit integer or retained in FP32, as well as the memory region in which it will be stored, according to the unit that consumes it. These decisions are sent as control parameters to the integrated kernel in the ACU through its AXI4-Lite control registers and determine the physical data layout. As illustrated using the \(K\) projection, the ACU produces its output in column-major order. Consequently, the quantized \(K\) heads assigned to the IMCU already appear as \(K^{\mathsf T}\) row by row and require no additional reordering. The FP32 \(K\) heads retained by the ACU pass through the integrated transposer before being written to memory, allowing the subsequent \(Q\times K^{\mathsf T}\) operation to read them directly in the required layout.
The crucial concern for CHARMING is whether processing quantized
scores from the IMCU creates a non-GEMM bottleneck, as these require
dequantization, softmax, and requantization, whereas FP32 scores
from the ACU require only softmax. To address this, we employ
lookup-table-based softmax for IMCU outputs and validate it on the Intel Xeon and VCK190 Arm Cortex-A72 (Table~\ref{tab:platform}). Figure~\ref{fig:softmax} confirms that processing the integer scores produced by the IMCU does not add latency to this non-GEMM stage.

\section{Results}
To evaluate the impact of CHARMING on transformer inference,
we examine the performance and energy benefits of partitioning
computation between IMCU and ACU with a focus on both the prefill and decode stages across various models with long sequences (Table~\ref{tab:models}) to observe the worst-case scenario on edge devices~\cite{yang2025qwen3,touvron2023Llama2,grattafiori2024Llama3,jiang2024mixtral,wan2025wan}. We first investigate how quantization affects the accuracy
and computational partitioning decisions underlying CHARMING.
\subsection{Quantization Study}

\subsubsection{Partitioning between ACU and IMCU}

\begin{table}[t]
\centering
\caption{Model dimensions used to evaluate CHARMING. $n$ is the sequence
length, $d$ the model dimension, $H$/$H_{kv}$ the query/key--value head
counts, $d_h$ the head dimension, $d_{\!f}$ the FFN dimension, and $L$ the
layer count. Every per-layer GEMM shape follows from these parameters.}
\label{tab:models}
\small
\setlength{\tabcolsep}{5pt}
\begin{tabular}{@{}l ccccccc@{}}
\toprule
\textbf{Model} & $n$ & $d$ & $H$ & $H_{kv}$ & $d_h$ & $d_{\!f}$ & $L$\\
\midrule
LLaMA 2 7B      & 8192  & 4096 & 32 & 32 & 128 & 11008 & 32\\
LLaMA 3 8B      & 8192  & 4096 & 32 & 8  & 128 & 14336 & 32\\
LLaMA 2 13B     & 8192  & 5120 & 40 & 40 & 128 & 13824 & 40\\
LLaMA 2/3 70B   & 8192  & 8192 & 64 & 8  & 128 & 28672 & 80\\
Mistral 7B v0.3 & 8192  & 4096 & 32 & 8  & 128 & 14336 & 32\\
Mixtral 8$\times$7B & 8192 & 4096 & 32 & 8 & 128 & 14336 & 32\\
WAN 2.2 A14B    & 32760 & 5120 & 40 & 40 & 128 & 13824 & 40\\
Qwen3-32B       & 32760 & 5120 & 64 & 8  & 128 & 25600 & 64\\
\bottomrule
\end{tabular}
\end{table}

\label{sec:partition}
\begin{figure}[t]
  \centering
  \includegraphics[width=\columnwidth]{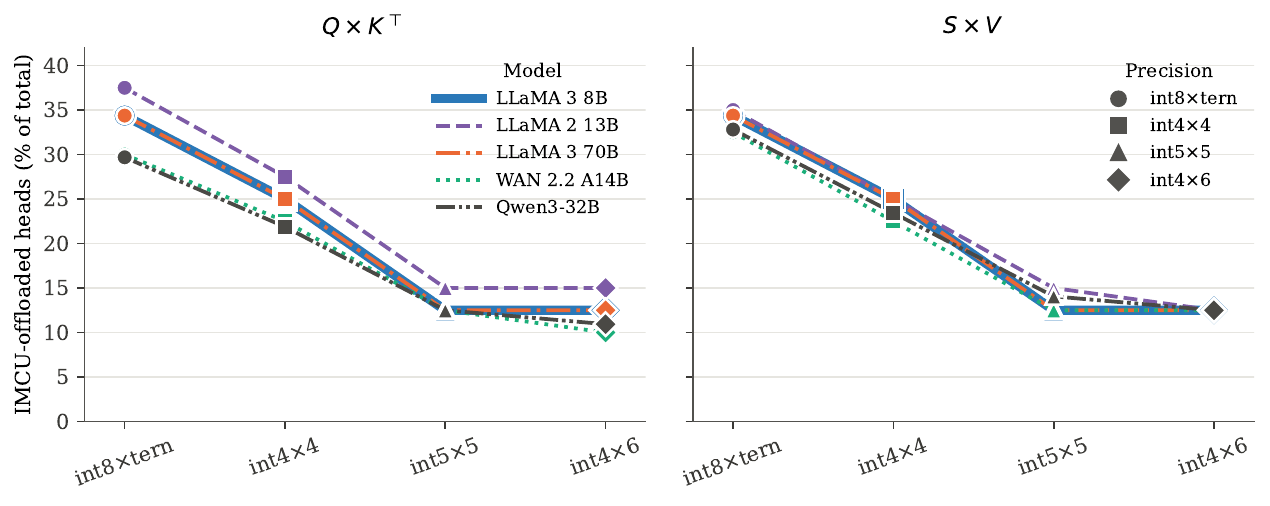}
  \vspace{-.2in}
  \caption{ IMCU-offloaded heads during prefill as a fraction of total heads, vs precision, for $Q\times K^{\top}$ (left) and $S\times V$ (right).}
  \vspace{-.1in}
  \label{fig:heads-split}
\end{figure}

CHARMING assigns each operation in a transformer layer to the substrate
that executes it most efficiently, and partitions only the operations
that benefit from both as shown in Figure~\ref{fig:dataflow}. The decision is driven by the in-memory latency
model of Equation~\ref{equ:latency} and Algorithm~\ref{alg:mapping}. As $T_{\mathrm{IMCU}}$ is
proportional to the reduction depth $K$ and output rows $M$, In-memory execution is, therefore,
cheaper when a reduction is shallow and expensive when it is deep, which
cleanly separates the operations of a transformer layer into two classes.
The projections and the feed-forward network, consequently, stay on the AIE array, as for these operations, $K$ ranges from $4096$ to $28672$.


In addition, these projection operations
also move comparatively little data---each of the $Q$, $K$, and $V$
projections accounts for only a few hundred megabytes per layer, as illustrated in
Figure~\ref{fig:dm-models}. As a result, there is little data-bus traffic for
memory read/write operations. Consequently, high throughput and low memory traffic  enable CHARMING to keep the projections, the output projection, and the FFN
entirely on the AIE.

On the contrary, the two attention stages, $Q\times K^{\top}$ and $S\times V$, have the
opposite profile in terms of data movement from and to the memory. While projection requires a few hundred megabytes of data transfer from memory, these operations require more than 8 gigabytes (Figure~\ref{fig:dm-models}), making them perfect for in-memory compute, which does not need to read/write over the external data bus. Besides,  $Q\times K^{\top}$ has a reduction depth of only $K = 128$, so $T_{\mathrm{IMCU}}$ is
small. Similarly, utilization of transpose, i.e., $V^{\top}\times S^{\top}$, in the case of $S\times V$, provides lower output rows $M$. As a result, both operations are the best fit for CHARMING concurrent operations. Figure~\ref{fig:heads-split} outlines the split for various models as well as for different precisions.

The partition, however, behaves differently in the decode stage, where each step
processes a single token (batch~1, context $L = 8192$).
$Q \times K^{\top}$ becomes one matrix--vector product per head with
$M = 1$, $K = d_h = 128$ and $N = 8192$. Its output fills exactly one
DRAM row, so a head occupies a single bank and uses every column of it.
As a result, offloading all of them costs a single MVM time:
1.18--4.06\,ms across the six precisions, against 3.91\,ms \emph{per
head} on the AIE array. Therefore, we set $N_{opt}^{QK} = H$ for decode. This offload is mostly hidden behind the $V$~projection that the AIE array executes concurrently. $S\times V$ behaves in the opposite way. Issued in transposed form,
$V^{\top} \times S^{\top}$ has $M = d_h = 128$, $K = 8192$ and
$N = 1$: each row operation produces one useful column out of 8192
(0.012\% column utilization), while the reduction depth $K$ sets the
per-head latency to $T_{\mathrm{IMCU}}(M,K,N) =$ 111.9--343.7\,ms, i.e.,\ 22--67$\times$
the 5.14\,ms the AIE array needs. The untransposed form does not help:
with $N = d_h$ the utilization rises only to 1.6\% and the depth is
unchanged. Therefore, we set $N_{opt}^{SV} = 0$ for the decode stage.

\begin{figure*}[t]
  \centering
  \begin{minipage}[t]{0.39\textwidth}
    \centering
    \includegraphics[width=\linewidth,height=1.3in,keepaspectratio]{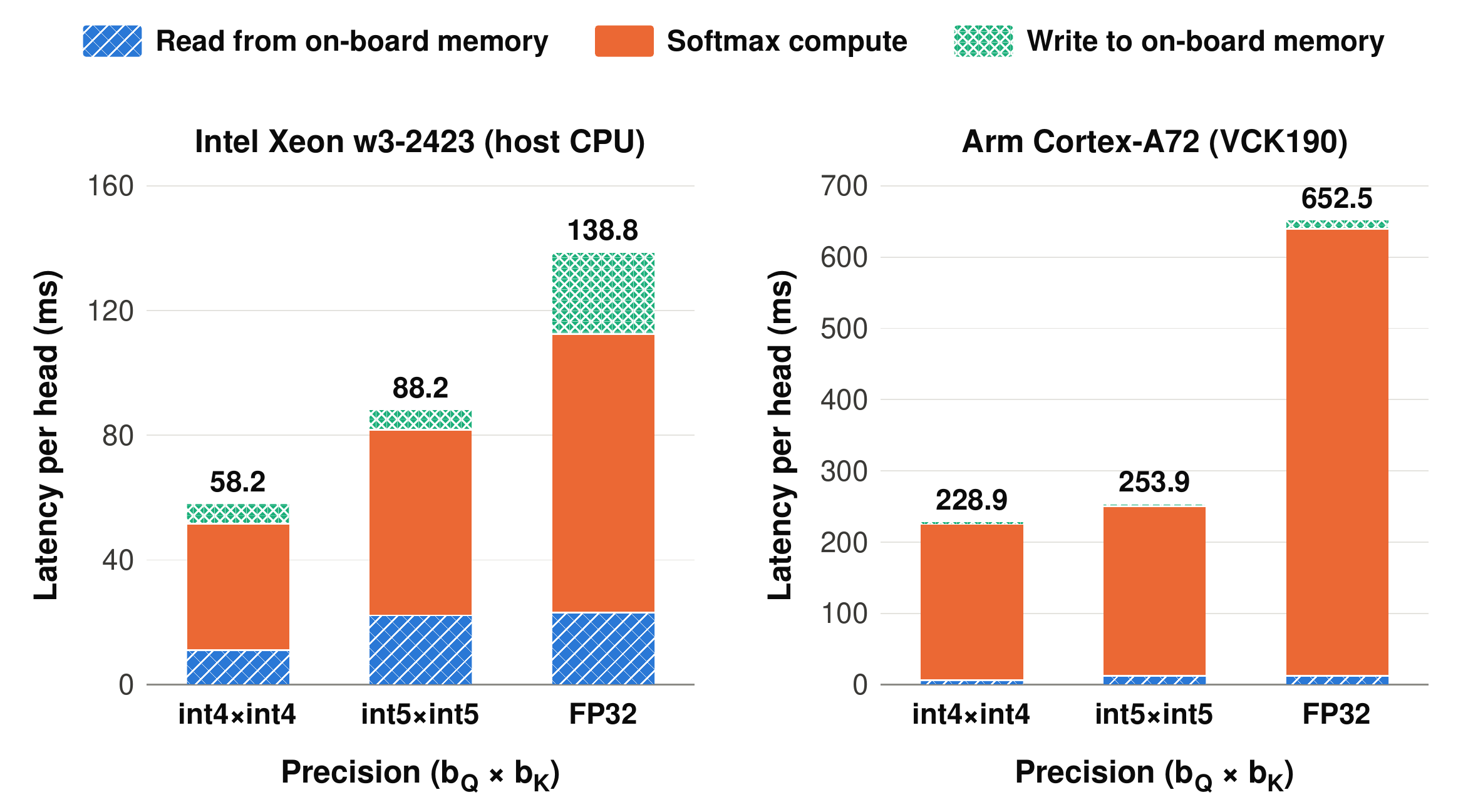}
    \vspace{-.1in}
    \caption{Softmax latency per \(8192\times8192\) attention head on the Intel Xeon w3-2423
    host CPU and on the Arm Cortex-A72 CPU on the VCK190.}
    \label{fig:softmax}
  \end{minipage}\hfill
  \begin{minipage}[t]{0.59\textwidth}
  \stepcounter{figure}
    \centering
    \begin{subfigure}[t]{0.3\linewidth}
      \centering
      \includegraphics[width=\linewidth,height=1.3in,keepaspectratio]{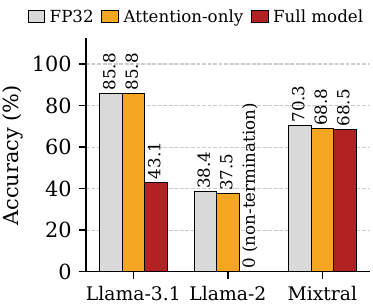}
      \caption{Quantization scope}
      \label{fig:quant-scope}
    \end{subfigure}\hfill
    \begin{subfigure}[t]{0.3\linewidth}
      \centering
      \includegraphics[width=\linewidth,height=1.3in,keepaspectratio]{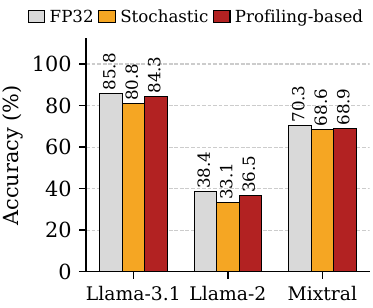}
      \caption{Head selection}
      \label{fig:head-selection}
    \end{subfigure}\hfill
    \begin{subfigure}[t]{0.4\linewidth}
      \centering
      \includegraphics[width=\linewidth,height=1.3in,keepaspectratio]{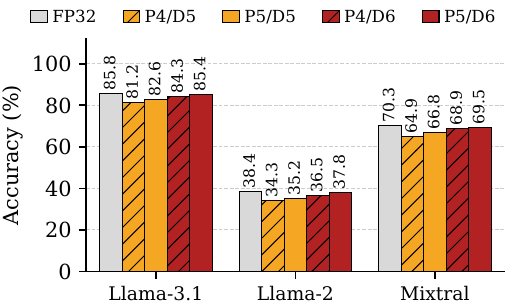}
      \caption{Quantization bit width}
      \label{fig:bit-width}
    \end{subfigure}
    \vspace{-.1in}
    \addtocounter{figure}{-1}
    \caption{Impact of quantization on accuracy.}
    \label{fig:quant}
  \end{minipage}
  \vspace{-.1in}
\end{figure*}

\subsubsection{Hardware Impact of Partitioning}
\begin{figure}[t]
  \centering
  \includegraphics[width=\linewidth]{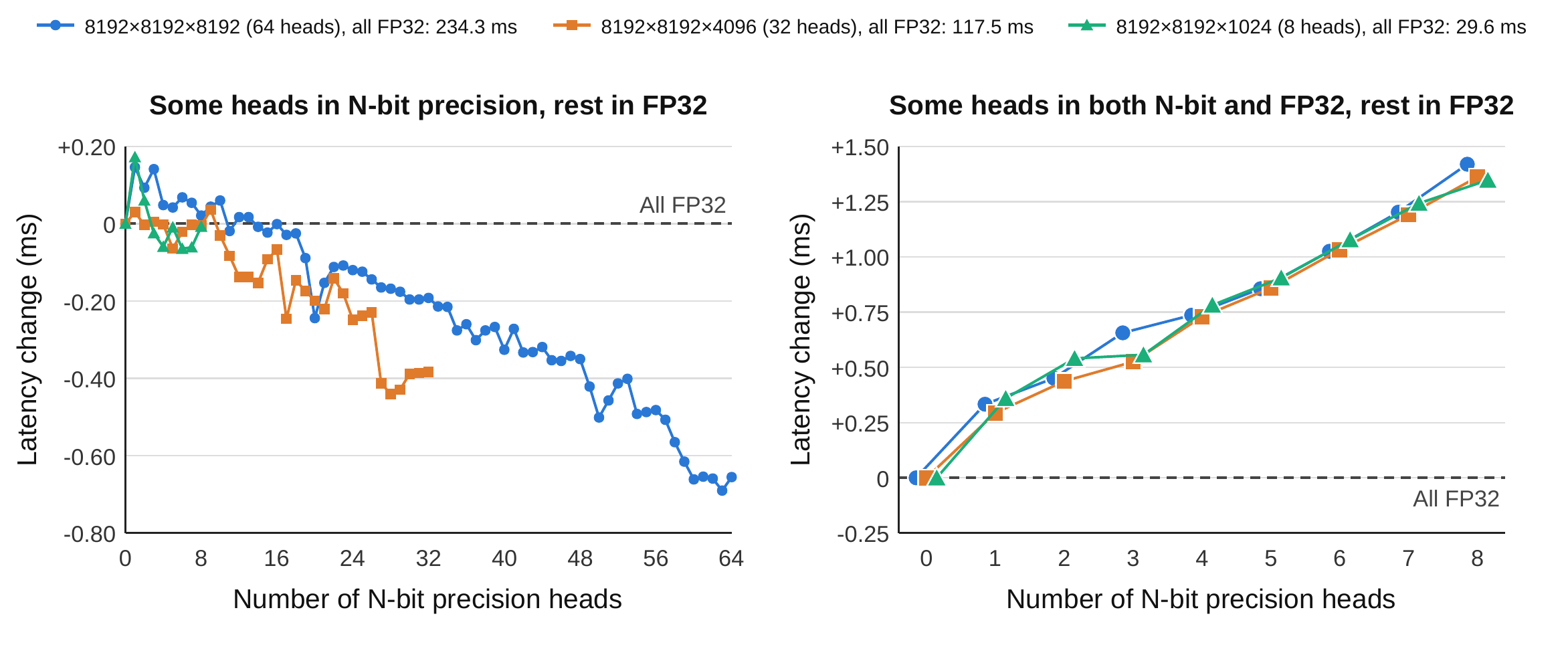}
  \caption{Latency change of the Q,K,V projection due to (left) some outputs in int8 and some in FP32 and (right) some outputs in FP32 and int8,both, and some in FP32 only.}
  \label{fig:partitioning-latency}
\end{figure}

Partitioning the attention heads requires three output mappings from the ACU projection kernel:
(i) FP32 heads for the ACU, written to memory region~0;
(ii) N-bit heads for the IMCU, written to memory region~1; and
(iii) with GQA, K/V heads whose query group is split between the ACU and the IMCU, are needed in both formats and are therefore written to both regions.
The compute latency itself is unchanged from CHARM~\cite{zhuang2024charm}; only the output write path in the PL differs. We measured the latency on the VCK5000 for three projection shapes, $8192\times8192\times N$ with $N = 8192$, $4096$ and $1024$ (64, 32 and 8 heads of dimension 128), as shown in Figure~\ref{fig:partitioning-latency}.
Moving heads to N-bit precision (ii) changes the latency by only $-0.69$ to $+0.17$~ms (below 0.6\%) across all head counts, because an N-bit head writes fewer bytes than an FP32 head.
A single shared head costs 0.29--0.36~ms, and eight shared heads, the most a GQA model with eight K/V heads can require, cost 1.35--1.42~ms. Partitioning therefore adds negligible hardware latency to the projection stage.

\subsubsection{Quantization Scope}
\label{sec:quantization-scope}
In this section, we compare the GSM8K accuracy of different quantization scopes. We define quantization scope as the set of operations executed in reduced precision. In CHARMING, PIM supports arbitrary bit width fixed point computation and AIEs support both fixed point computation (16 bits and 8 bits) and floating point computation (32 bits). We compare three different quantization scopes: no quantization (FP32 baseline), attention-only scope and full-model scope. Under the attention-only scope, only the attention score $Q\times K^{\top}$ and $S\times V$ computations are quantized. Under the full-model scope, the weights of all linear layers, including the QKV/output projections and the FFN layers, are additionally quantized to 8 bits. All runs use a generation budget of 1024 tokens, fixed across the baseline and every quantized configuration. The budget was set from the FP32 baseline's generation lengths (median 210 tokens, maximum 436), so 1024 leaves $2.35\times$ headroom. Quantization lengthens reasoning chains and can trigger degenerate repetition. The headroom costs nothing, since sequences that terminate normally stop at EOS and never spend it. A sequence that emits no terminator within the budget is scored incorrect.

Restricting the quantization scope to attention is effectively lossless across all three models: Llama-3.1-8B matches its FP32 baseline (85.8\% vs.\ 85.8\%), while Llama-2-13B and Mixtral-8x7B lose only 0.9 and 1.5 percentage points, respectively. Extending the same 8-bit quantization to the full model, however, causes severe degradation in both dense models: Llama-3.1-8B drops by 42.7 points to 43.1\%, and Llama-2-13B fails to emit an end-of-sequence token within the generation limit, yielding zero accuracy. Mixtral-8x7B is a notable exception, losing only 1.8 points under full-model quantization. We hypothesize that this robustness stems from its sparse Mixture-of-Experts (MoE) structure: because each expert processes only the tokens routed to it, activation quantization ranges are computed over smaller token subsets, which may limit the impact of outlier tokens; a detailed analysis is left to future work. Overall, in the dense models, the accuracy loss of 8-bit quantization originates from the non-attention layers rather than from attention itself. Attention can therefore be offloaded to low-precision PIM hardware without compromising model quality, which supports the quantization configuration of CHARMING, where the non-attention layers execute in FP32 on the ACU and attention executes in fixed point on the IMCU. The results are illustrated in Figure~\ref{fig:quant-scope}. We further examine whether this holds at the lower bit widths targeted by the IMCU in Section~\ref{sec:quantization-bitwidth}.

\subsubsection{Head Selection Mechanism} \label{sec:head-selection}
The search of Section~\ref{sec:mapping} fixes how many heads each stage offloads ($N^{QK}$, $N^{SV}$) and thus the latency; which heads it offloads determines the
accuracy. Here we evaluate two head selection mechanisms: profiling-based selection, which offloads the heads with the lowest damage scores (Algorithm~\ref{alg:mapping}, line~5), and stochastic selection. 
 
For a given subset size, stochastic selection randomly selects the same number of query heads in each layer. For profiling-based selection, we first quantize all heads and execute a forward pass on a small calibration set. We record a residual-weighted damage score for every head in each layer.
Let $o_h \in \mathbb{R}^{d_h}$ be the attention output of head $h$ at a given
token, taken before the output projection $W_O$, and let $\tilde{o}_h$ be the
same quantity with $Q$, $K$, $S$ and $V$ quantized at the target widths.
Since $W_O$ acts on the concatenated head outputs, its output decomposes as
$\sum_h W_O^{(h)} o_h$, where $W_O^{(h)} \in \mathbb{R}^{d \times d_h}$ is the
$h$-th column block of $W_O$. The error $e_h = \tilde{o}_h - o_h$ therefore
perturbs the residual stream by exactly $W_O^{(h)} e_h$. With $r$ the residual
stream leaving the layer and expectations taken over calibration tokens, the
damage score of head $h$ is
\begin{equation}
\mathrm{resid}_h
  = \frac{\mathbb{E}\!\left[\bigl\lVert W_O^{(h)} e_h \bigr\rVert_2^2\right]}
         {\mathbb{E}\!\left[\lVert r \rVert_2^2\right]} .
\label{eq:damage}
\end{equation}
Weighting by $W_O^{(h)}$ is what distinguishes this score from the local error
$\lVert e_h \rVert$: a head with a large local error but a small
output-projection block barely perturbs the residual stream.
Within each layer, we offload the lowest-scoring heads; under GQA, heads are
selected individually rather than by group (Section~\ref{sec:mapping}).
The denominator is constant within a layer and thus does not affect this
selection; it only expresses the score as a fraction of the residual-stream
energy, making scores comparable across layers for future cross-layer selection.
Each calibration set consists of only 4 problems randomly drawn from the GSM8K training set. Both mechanisms are repeated 5 times with different random draws or calibration sets, and we report the mean accuracy. We use 4-bit quantization in the prefill stage and 6-bit quantization in the decode stage. The results are reported in Fig.~\ref{fig:head-selection}. Stochastic selection reduces accuracy by 5.0 points on Llama-3.1-8B and 5.3 points on Llama-2-13B. Profiling-based selection limits these losses to 1.5 and 1.9 points, recovering roughly 70\% and 64\% of the accuracy lost under stochastic selection. On Mixtral-8x7B, both strategies remain within 1.7 points of the FP32 baseline, and they differ by only 0.3 points. We attribute this to the low sensitivity of Mixtral-8x7B to attention quantization at this bit-width setting: random selection already costs only 1.7 points, and the 0.3-point gap is within run-to-run variation, leaving little room for guided selection to improve upon random selection. Overall, profiling-based selection matches or outperforms stochastic selection on every model evaluated, and its advantage is greatest on the dense models, where the degradation is most severe, while requiring only four calibration problems.

\subsubsection{Prefill vs. Decode Quantization}
\label{sec:quantization-bitwidth}
We further study the effect of assigning different bit widths to the prefill and decode phases, evaluating prefill at 4 and 5 bits and decode at 5 and 6 bits with profiling-based selection. As illustrated in Figure~\ref{fig:bit-width}, across all three models, decode precision has a larger impact on accuracy than prefill precision. Raising decode from 5 to 6 bits improves accuracy by 2.2--3.9 points, whereas raising prefill from 4 to 5 bits yields only 0.6--1.9 points. As a result, the 4-bit prefill / 6-bit decode configuration consistently outperforms 5-bit prefill / 5-bit decode, even though both use the same total bit budget (84.3 vs. 82.6 on Llama-3.1-8B, 36.5 vs. 35.2 on Llama-2-13B, and 68.9 vs. 66.8 on Mixtral-8x7B). We attribute this asymmetry to error accumulation in autoregressive decoding: decode attention is executed once per generated token, and GSM8K requires long chain-of-thought outputs, so quantization errors introduced during decoding compound over the generated sequence, while prefill errors are incurred only once over the prompt. With 5-bit prefill and 6-bit decode, all three models remain within 0.9 points of their FP32 baselines. These results suggest that, under a fixed precision budget, bits are better spent on the decode phase than on the prefill phase.

\subsection{CHARMING Evaluation}
The preceding quantization study demonstrates that, with
appropriate partitioning and attention-head selection,
CHARMING can preserve inference accuracy while retaining
FP32 computation on the AIEs and executing selected attention
heads at substantially lower fixed-point precision.
CHARMING exploits this flexibility by offloading these
operations to in-memory processing, reducing data movement
and enabling concurrent execution with the AIEs.
We next evaluate the resulting performance and energy benefits
during prefill and decode.


\subsubsection{Performance Improvement}

\begin{figure*}[t]
  \centering
  \includegraphics[width=\textwidth]{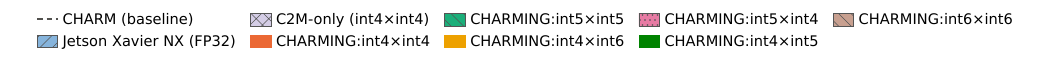}\\[-2pt]
  \begin{subfigure}[t]{0.325\textwidth}
    \centering
    \includegraphics[width=\linewidth]{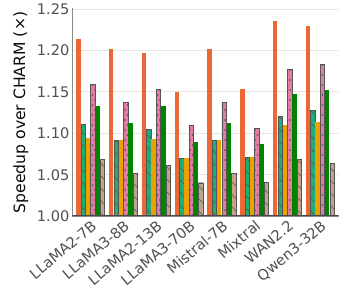}
    \vspace{-.3in}
    \caption{Speedup of CHARMING over CHARM.}
    \label{fig:speedup}
  \end{subfigure}\hfill
  \begin{subfigure}[t]{0.325\textwidth}
    \centering
    \includegraphics[width=\linewidth]{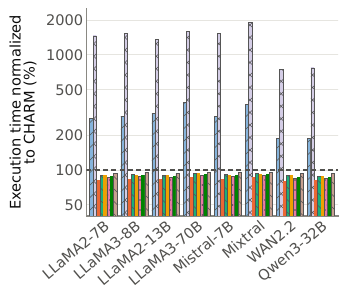}
    \vspace{-.3in}
    \caption{Execution time normalized to CHARM.}
    \label{fig:normalized-time}
  \end{subfigure}\hfill
  \begin{subfigure}[t]{0.325\textwidth}
    \centering
    \includegraphics[width=\linewidth]{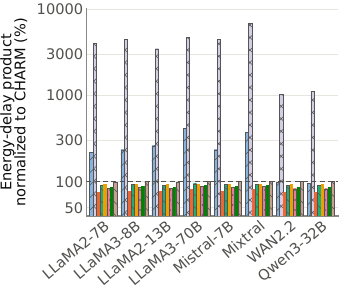}
    \vspace{-.3in}
    \caption{Energy-delay product normalized to CHARM.}
    \label{fig:edp}
  \end{subfigure}
  \vspace{-.1in}
  \caption{Prefill performance and energy-delay product of CHARMING at every precision, compared with
  CHARM (dashed baseline), C2M-only and the Jetson Xavier NX.}
  \label{fig:performance-edp}
  \vspace{-.1in}
\end{figure*}

\begin{figure*}[t]
  \centering
  \begin{subfigure}[t]{0.32\textwidth}
    \centering
    \includegraphics[width=\linewidth]{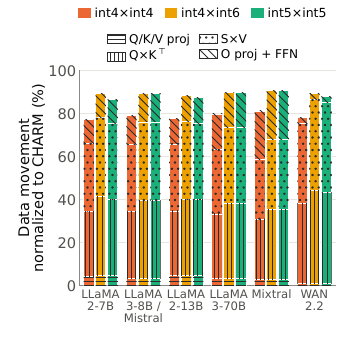}
    \vspace{-.3in}
    \caption{Data movement per layer.}
    \label{fig:dm-models}
  \end{subfigure}\hfill
  \begin{subfigure}[t]{0.36\textwidth}
    \centering
    \includegraphics[width=\linewidth]{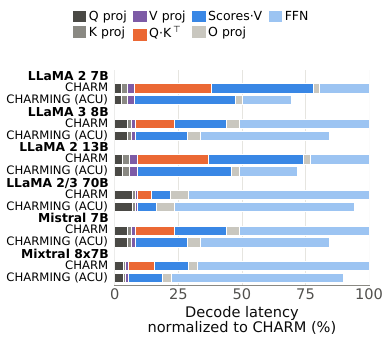}
    \vspace{-.3in}
    \caption{Decode latency per layer.}
    \label{fig:decode-latency}
  \end{subfigure}\hfill
  \begin{subfigure}[t]{0.31\textwidth}
    \centering
    \includegraphics[width=\linewidth]{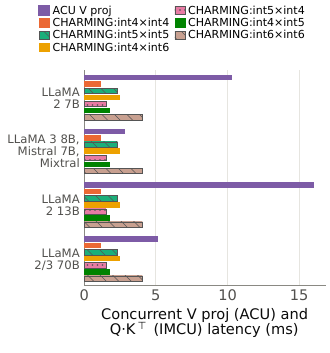}
    \vspace{-.3in}
    \caption{Concurrent ACU and IMCU in decode.}
    \label{fig:decode-concurrent}
  \end{subfigure}
  \vspace{-.1in}
  \caption{Data movement and concurrent execution in CHARMING.
  }
  \vspace{-.1in}
  \label{fig:dm_reduction}
\end{figure*}

Figures~\ref{fig:speedup} and~\ref{fig:normalized-time}
report per-layer execution time under CHARM and
CHARMING, normalized to the CHARM baseline.
As illustrated in Figure~\ref{fig:dataflow}, because
the ACU and IMCU execute concurrently, the latency of
each attention operation is determined by the longer
of their execution times. Offloading heads to the IMCU
therefore reduces latency when it decreases the maximum
execution time of the two units, allowing subsequent
operations to begin earlier. CHARMING thus reduces
overall execution time without altering the projections,
softmax, or FFN.

Among the evaluated configurations, $\mathrm{int_4}\times\mathrm{int_4}$
delivers the largest speedup, as its low per-head in-memory cost allows
the most heads to be offloaded before the parallel execution balances. For LLaMA~2~7B
it reduces per-layer time from $2{,}957$~ms to $2{,}437$~ms, a
$1.21\times$ speedup ($17.6\%$); the higher-precision configurations
offload fewer heads and yield smaller gains, from $1.07\times$ at
$\mathrm{int6}\times\mathrm{int6}$ to $1.16\times$ at
$\mathrm{int5}\times\mathrm{int4}$.

The benefit grows with a model's attention share. Short-sequence LLMs
such as LLaMA and Mistral, where attention is $52$--$69\%$ of the layer,
achieve $13$--$18\%$ reductions ($1.15$--$1.21\times$), whereas the long-sequence models WAN~2.2
and Qwen3-32B, whose $n = 32{,}760$ makes attention $86$--$87\%$ of the layer,
reach up to $19.1\%$ ($1.24\times$). Because the offloaded attention grows
quadratically with sequence length while the projection and FFN cost grows
linearly, CHARMING scales favorably to the long-context regime that most
stresses conventional accelerators.

We also benchmarked the execution time of C2M with HBM2, which we use for the IMCU in CHARMING, as well as
the Jetson Xavier NX. Comparing with C2M and CHARM indicates the significance of CHARMING, as it improves
the performance significantly across various models while leveraging the benefits of both CHARM and C2M
by reducing data movement and improving execution time: running every GEMM in memory (C2M-only) is $9.2$--$22.1\times$ slower than CHARMING. CHARMING, in addition, outperforms the GPU-based Jetson
Xavier NX, as its per-layer latency is $131$--$347\%$ higher than CHARMING.

The performance improvements arise from two complementary effects:
reducing external data movement for attention heads assigned to the
IMCU and overlapping in-memory computation with ACU execution.

Unlike architectures where PIM and conventional memory accesses
compete for the same DRAM banks, CHARMING assigns the ACU and
IMCU to separate physical memory regions. This allows both units to
exploit bank-level parallelism independently, avoiding memory
contention during concurrent execution.

The ACU transfers operands from memory through the PL
to the AIE and returns results through the PL for accumulation
in URAM (Figure~\ref{fig:architecture}). In contrast, the IMCU
executes operations in place, eliminating external operand transfers
for the offloaded attention heads. Consequently, CHARMING reduces
data movement per layer by 9.4--22.8\% across models and precisions,
corresponding to 2.2--8.9\,GB for models with 8K tokens and
39--126\,GB for WAN~2.2 and Qwen3 with 32K tokens
(Figure~\ref{fig:dm-models}).

During decode, the benefits of in-memory execution extend beyond
data movement reduction to concurrent execution. As shown in
Figure~\ref{fig:decode-latency}, CHARMING achieves up to 44\% speedup over CHARM
by overlapping IMCU computation with ACU execution.
While the ACU executes the $V$ projection in 2.88--16.02\,ms,
the IMCU completes $Q\times K^{\mathsf T}$ in 1.18--4.06\,ms,
with its execution mostly hidden  behind the ACU's $V$
projection (Figure~\ref{fig:decode-concurrent}). Because the
two units operate on independent memories, IMCU execution
does not introduce bank-level contention with the ACU's
external memory accesses.

\subsubsection{Energy Characterization}
\label{energy_characterization}

To measure the energy, we used a direct measurement approach. We measured the total power of the VCK5000
board while running each model, which gives the ACU energy including the board's idle
power. For the IMCU, we used the energy reported by gem5-CIMD for each in-memory operation. We estimate CHARMING's total energy by combining the
measured VCK5000 board energy with the simulated IMCU
energy.
Figure~\ref{fig:edp} reports the energy-delay product (EDP), which captures the two crucial costs of a
system, energy and time, in one metric.

Compared with CHARM, CHARMING:int4$\times$int4 saves up to $7.6\%$ energy ($5.5$--$7.6\%$) during prefill
and lowers EDP by $17.8$--$24.7\%$. The energy saving comes from the offloaded heads: an attention head
costs $0.78$~J ($Q\times K^{\top}$) and $1.15$~J ($S\times V$) in memory, against $1.27$~J and $2.14$~J
on the AIE. EDP, therefore, improves about three times more because CHARMING reduces energy and
execution time together (for LLaMA~2~7B, $0.924\times0.824 = 0.761$ of CHARM). During decode, CHARMING
saves $4.9$--$27.9\%$ energy and lowers EDP by $10.3$--$49.8\%$. Each decode head on the AIE is a small,
launch-overhead-bound GEMV, whereas the IMCU computes all $Q\times K^{\top}$ heads in a single round that is
hidden behind the V projection. The gain is therefore largest for multi-head models with many attention
GEMVs per layer (LLaMA~2~7B and 13B: $26$--$28\%$) and smallest for grouped-query models with only eight
KV heads (LLaMA~3~70B: $4.9\%$).

Compared with the Jetson Xavier NX, CHARMING lowers EDP by $21.2$--$80.1\%$ in prefill. The Jetson draws
only about $15$~W and therefore consumes less energy per layer for most models, but it is
$2.3$--$4.5\times$ slower than CHARMING, and EDP accounts for this delay. The advantage is largest for the
70B models, whose large GEMMs keep the AIE array most efficient, and smallest for the long-sequence
models WAN~2.2 and Qwen3-32B.

We also estimated the energy associated with external data movement
by continuously reading operands from DDR into the PL. At approximately
2.5~nJ per byte transferred, CHARMING avoids an estimated 340~mJ of data transfer energy per layer
by eliminating 135~MB of $Q\times K^{\mathsf T}$ data movement
during decode for LLaMA~2~7B. Although the IMCU incurs its own
computation energy, this eliminates the corresponding data movement
energy for attention heads offloaded to the IMCU.

\section{Conclusion}
\label{sec:conclusion}

This work presents CHARMING, a heterogeneous acceleration
approach that exploits the complementary capabilities of
tensor processing (ACU) and in-memory computing (IMCU) to improve the
performance and energy efficiency of transformer inference.
By partitioning attention computation across these substrates,
CHARMING reduces data movement and overlaps computation
without requiring the entire workload to execute at the
precision or performance characteristics of either substrate.

A key contribution of CHARMING is its accuracy- and
hardware-aware partitioning strategy. Our quantization study
demonstrates that selected attention heads can execute at
substantially reduced fixed-point precision while the
remaining computation retains FP32 precision on the AIEs,
preserving model accuracy through appropriate head selection.
This flexibility enables CHARMING to adapt the partition
to model characteristics, sequence length, and quantization
precision, balancing execution time across the two substrates.

Our evaluation demonstrates that CHARMING improves per-layer
prefill performance by up to 1.24$\times$ and decode performance by up to 1.44$\times$ relative to CHARM.
By eliminating external operand transfers for offloaded
attention heads, CHARMING reduces data movement by
9.4--22.8\% and improves the energy-delay product by up to
24.7\% during prefill and 49.8\% during decode.
These results demonstrate the potential of combining
programmable tensor acceleration with processing-in-memory
to address the growing computational and data movement
demands of transformer inference.


\begin{acks}
\end{acks}

\bibliographystyle{ACM-Reference-Format}
\bibliography{software}

\end{document}